\documentclass[twocolumn,english,prx]{revtex4-1}

\usepackage{graphicx}
\usepackage{amssymb}

\begin{document}

\author{Bart\l{}omiej Szafran}
\author{Alina Mre\'nca-Kolasi\'nska}
\author{Dariusz \.Zebrowski}

\address{AGH University of Science and Technology, Faculty of Physics and Applied Computer Science, al. Mickiewicza 30, 30-059 Krak\'ow}



\title{Finite difference method for Dirac electrons in circular quantum dots}



\begin{abstract}
A simple and reliable finite difference approach is presented for solution of the Dirac equation eigenproblem
for states confined in rotationally symmetric systems. The method sets the boundary condition for the spinor wave function components at the external edge of the system
and then sweeps the radial mesh in search  for the energies for which the boundary conditions are met inside the flake. 
The sweep that is performed from the edge of the system towards the origin allows for application of a two-point finite difference quotient
of the first derivative, which prevents the fermion doubling problem and the appearance of the spurious solutions with rapid oscillations of the wave functions.
\end{abstract}
\maketitle




\section{Introduction}

Numerical solutions of the Dirac equation for relativistic particles 
have been performed since the 1970's in the context of the problems of nuclear physics
and the lattice gauge theories \cite{kogut}. The interest in computational approaches
for the massless Dirac equation (Weyl equation) has been extended to the solid state with the arrival of graphene \cite{neto}, the two-dimensional material with  
gapless and linear dispersion relation near the charge neutrality point. 

Solution of the discretized version of the Dirac Hamiltonian on a mesh is commonly pested with 
spurious solutions that are characterized by rapidly oscillating wave functions,  large momenta
but low expectation values of the energy \cite{kogut,stacey,bender,fermdi,poszl}. The spurious solutions
are degenerate with the regular ones which results in the so-called fermion doubling problem \cite{kogut,stacey,tanimura,bottcher,wilson,susskind,tworzydlo,spl4}.
The problem arises in particular with the central, three-point discretization of the first order spatial derivative in the Dirac
Hamiltonian. The central finite difference quotient misses the wave function oscillations that occur with the periodicity of the mesh spacing  \cite{stacey,bender,bottcher,tworzydlo}.  
The spurious solutions are also observed in the finite element method \cite{poszl} despite the exact treatment
of shape functions derivatives in this approach. The spurious states can be removed with Wilson approach \cite{stacey,tanimura,wilson,susskind} that introduces artificial terms
in the Hamiltonian depending on the square of the electron momentum, which removes the spurious states from the low-energy spectrum.
Alternatively, the shift of the finite difference quotients can be applied which allows the lattice derivative
to resolve the rapid oscillations of the wave function \cite{stacey,bottcher,bender,tworzydlo}.

In this paper, we focus on confined solutions of the Dirac type effective Hamiltonians for graphene. The Weyl fermions evade the confinement 
by the electrostatic potentials with the Klein tunneling phenomenon \cite{k1,k2,k3}.
 A way to confine the carriers is to either use a finite
flake of graphene \cite{gru,zarenia,flake1,flake2,flake4,flake5,mori,Yamamoto,Zhang,Guclu,Ezawaf,Potasz1,tho} or
 introduce a non-zero mass in the region outside of the dot \cite{berry,nori}. The non-zero mass along with a
finite energy gap is introduced to the effective Hamiltonian by the influence of lattice-matched substrates \cite{sachs}. 
The gap or non-zero mass can be locally introduced to silicene, a graphene-like 2D hexagonal Si crystal with buckled crystal lattice,
using  electric fields vertical to the surface  \cite{ni,Drummond12}. 

In this paper we present a very simple and effective finite difference method for determination of the Dirac Hamiltonian eigenstates localized in circular quantum dots,
applicable to both the finite flakes and systems with spatially modulated energy gap. The method is based on a two-point backward derivative. Starting from
the boundary condition at the external edge of the flake for the two components of the eigenfunctions one can pin the energies for which the 
boundary conditions in the interior of the flake are fulfilled for both the spinor components. The proposed mesh-sweeping method resolves only the actual solutions
and is free of the spurious ones, hence extra terms in the Hamiltonian of a Wilson type or a post-treatment of the eigenstates are not necessary. 

The paper is organized as follows. In Section II we present the Hamiltonian. The analytical solution given in Section III is used for
the test calculations. In Section IV we illustrate the problem of the spurious solutions, the fermion doubling and the Wilson procedure using a diagonalization
of the finite difference Hamiltonian with the central three point lattice derivative.  The original  method reported in this work is presented in Section V. Examples of applications, including
the spatial variation of the energy gap and a graphene quantum ring are presented in Section VI. The summary is given in Section VII.

\section{Hamiltonian}
The effective low-energy Dirac Hamiltonian for electrons in graphene or silicene can be separated into two $2\times 2$ operators
each associated with one of the non-equivalent valleys of the Brillouin zone, $K$ or $K'$. These valleys  will be referred 
to by the  index $\eta=1$ and $-1$, respectively. 
The $2\times 2$ Hamiltonian for the valley index $\eta$  takes the form \cite{neto}
\begin{equation}
H_\eta=\left(\begin{array}{cc} U_A({\bf r}) & \hbar v_f (k_x -i \eta k_y) \\ \hbar v_f ( k_x +i \eta k_y) & U_B({\bf r})\end{array}\right), \label{operator}
\end{equation}
where $v_f$ is the Fermi velocity, ${\bf k}=-i\nabla +\frac{e}{\hbar}{\bf A}$, and ${\bf A}$ is the vector potential.
This Hamiltonian acts on a wave function of form $\Psi({\bf r})=\left(\begin{array}{c}\Psi_{1}({\bf r}) \\ \Psi_{2}({\bf r}) \end{array}\right)$,
whose components correspond to the A and B graphene sublattices \cite{neto}, respectively. In Eq. (\ref{operator}), 
 $U_A$ and $U_B$ stand for potentials at the two sublattices.
 The potentials can be made unequal 
due to the substrate of e.g.  a hexagonal boron nitride \cite{sachs}. Vertical electric field applied to a graphene-like 2D hexagonal crystal with the buckled crystal lattice, e.g. the silicene 
 also introduces unequal $U_A$ and $U_B$ values, that introduces the energy gap to the dispersion relation \cite{ni,Drummond12}.

Hamiltonian for a circular flake of graphene in the vertical magnetic field $(0,0,B)$ and symmetric gauge ${\bf A}=(-By/2,Bx/2,0)$ commutes with
generalized angular momentum operator $J_z=L_z{\bf I}+\eta \frac{\hbar }{2}\sigma_z$, where $L_z$ is the 
z-component of the orbital angular momentum. For the polar coordinates: $r=\sqrt{x^2+y^2}$ and $\phi=\arctan(y/x)$, the operator is  $L_z=-i\hbar \frac{\partial}{\partial \phi}$ and $\sigma_z$ is the Pauli matrix in the sublattice space.
Therefore, the stationary states can be labeled with magnetic quantum number $m$ and have form
 \begin{equation}
\Psi_{m,\eta}=\exp(im\phi) \left(\begin{array}{c} f_1 (r) \\ f_2 (r) \exp(i\eta \phi) \end{array} \right). \label{laal}
\end{equation}
The radial functions $f_1(r)$ and $f_2(r)$ of Eq. (\ref{laal}) are solutions to the system of eigenequations
\begin{equation}
\left\{ \begin{array}{c}  U_A(r)f_1+v_F\left[-\eta\frac{i\hbar}{r}(m+\eta)f_2-i\hbar f'_2-\eta \frac{ieBr}{2}f_2 \right]=E f_1,\label{e1}\\
 U_B(r)f_2+v_F\left[ \eta\frac{i\hbar}{r}m f_1-i\hbar f'_1+\eta \frac{ieBr}{2}f_1 \right]=E f_2.  \end{array}\right.
\end{equation}

\section{Analytical solution}

An analytical solution \cite{gru,tho} to Eq. ($\ref{e1}$)  without the external fields 
will be used for the test calculations. For $B=0$ and $U_A=U_B=0$, Eq. ($\ref{e1}$) reads
\begin{eqnarray}
&& v_F\left[-\eta\frac{i\hbar}{r}(m+\eta)f_2-i\hbar f'_2\right]=E f_1, \label{er1}\\
&& v_F\left[ \eta\frac{i\hbar}{r}m f_1-i\hbar f'_1 \right]=E f_2. \label{er2}
\end{eqnarray}
For $E\neq 0$ one can eliminate $f_2$ from (\ref{er2}) and plug it into (\ref{er1}) to obtain 
\begin{equation}
r^2 f_1''+rf_1'+\left(\frac{r^2E^2}{v_f^2\hbar^2}-m^2\right)f_1=0,
\end{equation}
which upon introduction of a dimensionless radial coordinate $\rho=\frac{Er}{\hbar v_f}$,
produces the Bessel equation 
\begin{equation}
\rho^2 f_1''+\rho f_1'+(\rho^2-m^2)f_1=0,
\end{equation}
where, up to a normalization constant $f_1=J_m(\rho)$, and $J_m$ is the $m$th Bessel function of the first kind.
Using identities \cite{bee} $\frac{2}{\rho}J_m=J_{m-1}+J_{m+1}$, $2J_m'=J_{m-1}-J_{m+1}$,
one finds $f_2=iJ_{m+\eta}(\rho)$. 
For the test calculations we take a finite flake of radius $R$ and apply a so-called zigzag boundary condition \cite{gru}. For the flake that is terminated by the
$A$ sublattice, the second component of the wave function -- that describes the $B$ sublattice, needs to vanish at $r=R$, hence
$f_2(\frac{E_{mn} R}{v_f\hbar})=0$. This condition defines the spectrum of the states confined within the flake, where
 $n$ numbers the eigenvalues for fixed $m$. For the conduction band states
we use positive values of $n$. From the boundary condition we find the non-zero energy eigenvalue 
$$ E_{mn}=\mathrm{sign}(n)\frac{Z_{{|m+\eta|},|n|} v_f \hbar }{R}$$
where $Z_{|{m+\eta}|,|n|}$ is the $n$th zero of $J_{|m+\eta|}$ Bessel function \cite{bee}  -- see Table \ref{besle}.

\begin{table}
\begin{tabular} {c|ccc}
 & $n=1$ & $n=2$ &  $n=3$  \\ \hline
$m=-2$ & 3.831706 & 7.015587 & 10.173468 \\ $m=-1$ & 2.404825 & 5.520078 & 8.653727 \\
$m=0$ & 3.831706 & 7.015587 &  10.173468 \\ $m=1$ & 5.135622 & 8.417244 & 11.619841 \\
\end{tabular}
\caption{Zeroes of the $J_{|m+\eta|}$ Bessel function for $\eta=1$, or energies of states confined within a circular flake of radius $R$ and zigzag termination in the units of $\frac{R}{v_f \hbar}$.} \label{besle}
\end{table}

\section{Diagonalization of the finite difference problem}
For illustration of the problem of spurious solutions, 
we solve Eqs. (\ref{er1},\ref{er2}) in a finite difference approach using diagonalization 
of the resulting algebraic  equations. For that purpose we need a boundary
condition at the origin. 
The radial functions $f_1$, $f_2$, near $r=0$ behave as $r^{|m|}$ and $r^{|m+\eta|}$, respectively. 
A natural Dirichlet  condition at $r=0$ independent of the orbital angular momentum can be obtained by substitution $f_i=\frac{\psi_i}{r}$ for $i=1,2$. Since $f_i$ is finite at $r=0$, 
$\psi_i$ needs to vanish at the origin independent of $m$. The system of eigenequations for  functions $\psi_i$ then reads
\begin{eqnarray}
& & v_F\left[- \frac{i\hbar \eta m}{r}\psi_2-i\hbar \psi'_2\right]=E \psi_1, \label{era1}\\
&& v_F\left[ \frac{i\hbar(\eta m +1)}{r}\psi_1-i\hbar \psi'_1 \right]=E \psi_2. \label{era2}
\end{eqnarray}
The boundary conditions are then $\psi_1(0)=\psi_2(0)=0$ at the origin and the zigzag boundary at the edge $\psi_1(R)=1$, $\psi_2(R)=0$.
We use $N+1$ points on the radial mesh, including the origin $r=0$ where the boundary conditions are applied,  and the discretization step $dr=\frac{R}{N}$.
For discretization we use the difference quotient $\psi'=\frac{\psi(r+dr)-\psi(r-dr)}{2dr}$ with the exception of the last point on the mesh for $r=R$,
where a two point formula is used $\psi(R)=\frac{\psi(R)-\psi(R-dr)}{dr}$. The resulting algebraic eigenequation
is described by a non-symmetric matrix and is solved with the ARPACK library (procedure ZGEEV). 

For the numerical calculations we use $v_f=\frac{3ta}{2\hbar}$, 
for the silicene tight-binding hopping energy of $t=1.6$ eV and the nearest
neighbour distance of  $a=0.225$ nm. 
For the radius of the flake we take $R=80$ nm.  The positive eigenenergies are listed in Table \ref{pozytywne} for $m=0$, and $\eta=1$ and a varied number of points on the spatial mesh. 
In Table \ref{pozytywne} we number the eigenstates by $n'$ -- the number that counts
also the spurious states -- to distinguish from the actual quantum number $n$ used in Table \ref{besle} and below.
For $m=0$ and $\eta=1$ one would expect to obtain the values given by $n'=2$ and $n'=4$ -- see Table \ref{besle}.  The other values are the energies of the spurious
states. In fact, the energies of spurious solutions obtained for $m=0$ correspond to the regular eigenstates but for a different $m$. 
In particular, the energies for odd $n'$ of Table \ref{pozytywne} that were calculated with $m=0$ correspond to the exact energies for $m=-1$ -- see Table \ref{besle}.
With the spurious solutions the degeneracy of energy levels is artificially increased by a factor of two, which produces the fermion doubling problem  \cite{kogut,stacey,tanimura,bottcher,wilson,susskind,tworzydlo,spl4}.

\begin{figure}[htbp]
\begin{tabular} {cc}
 \includegraphics[width=0.475\columnwidth]{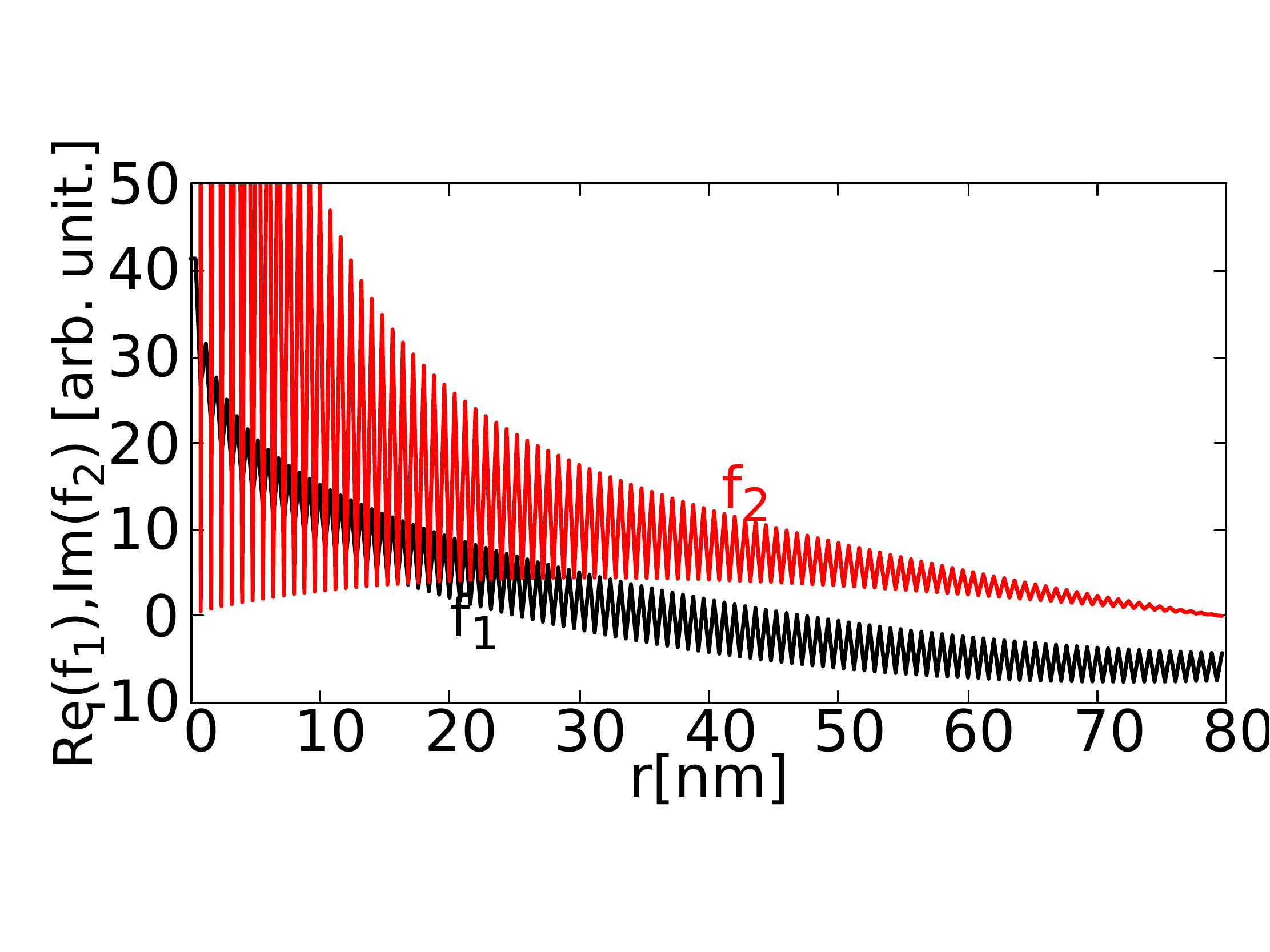} & \includegraphics[width=0.475\columnwidth]{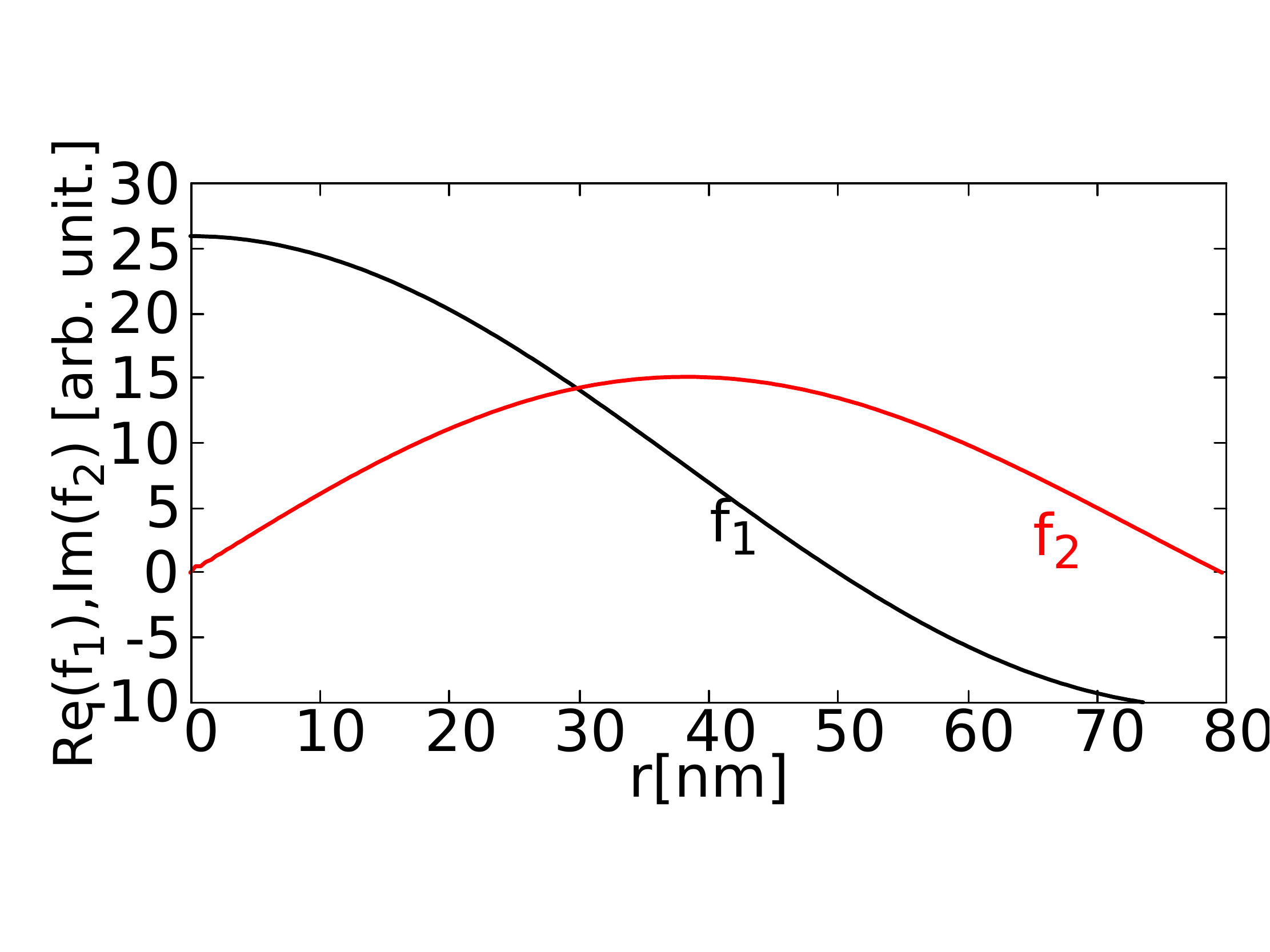}\\
 \includegraphics[width=0.475\columnwidth]{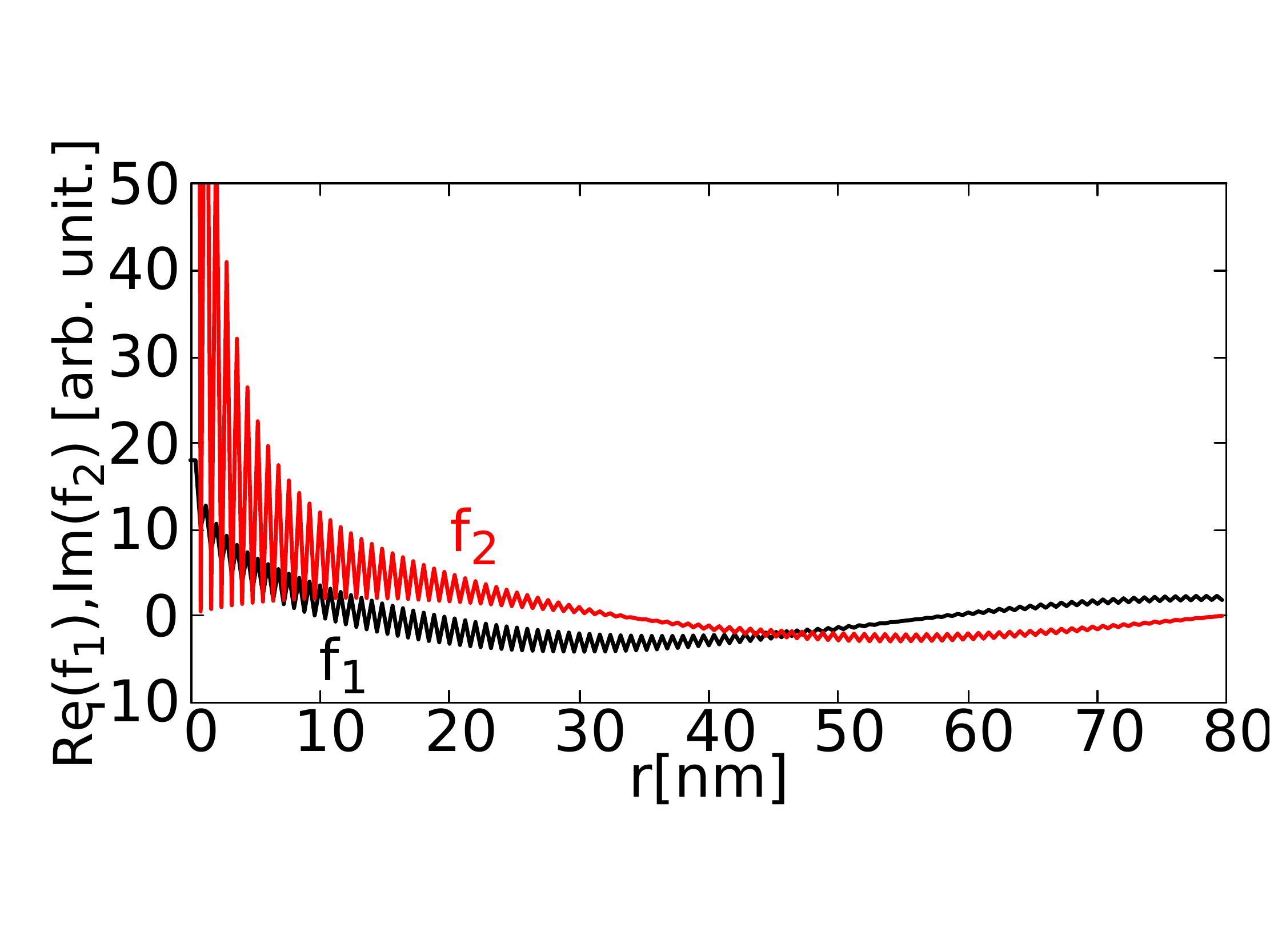} & \includegraphics[width=0.475\columnwidth]{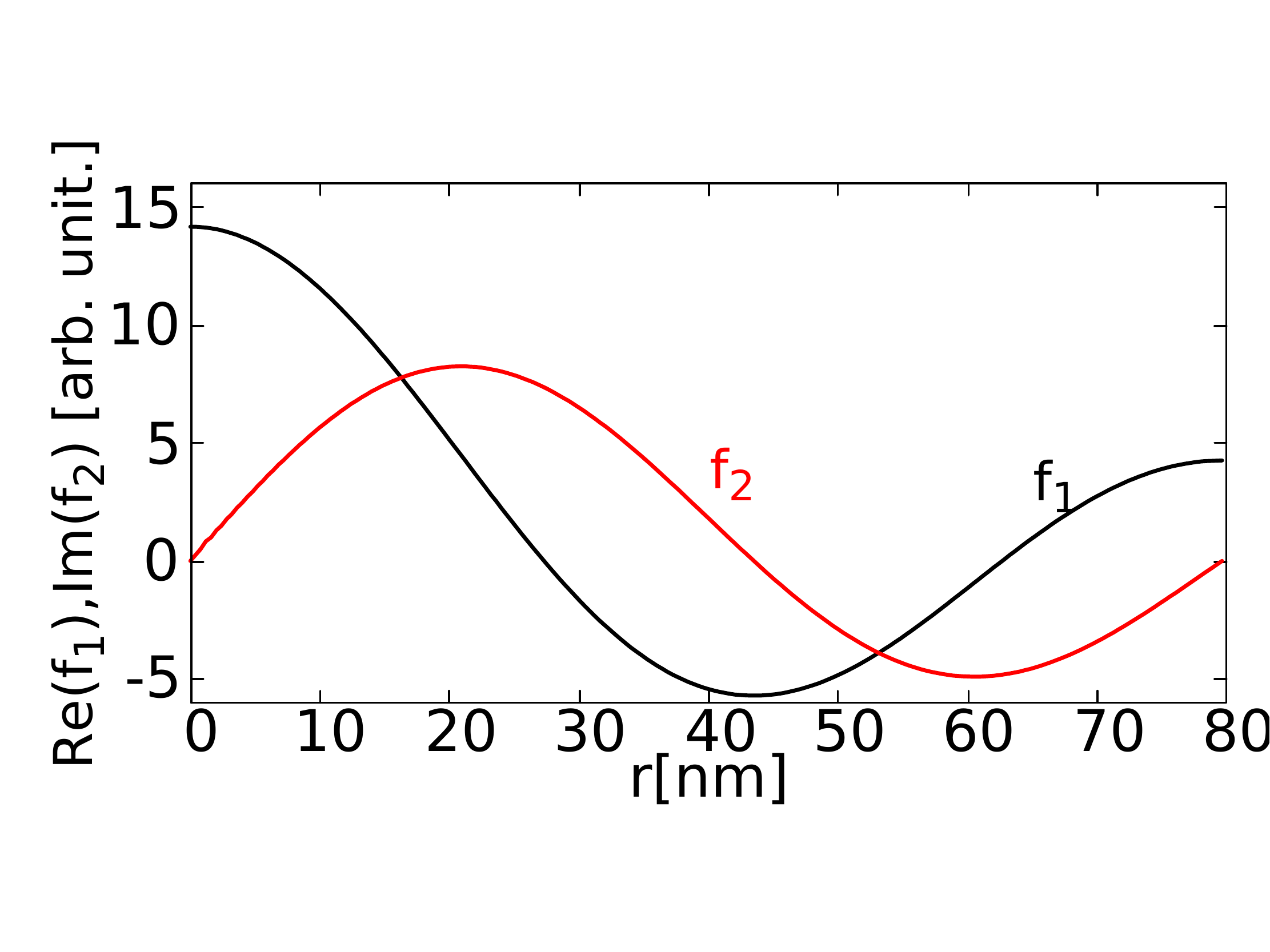}
\end{tabular}
 \put(-150,60){(a)} \put(-150,-30){(c)} \put(-20,60){(b)} \put(-20,-30){(d)}
\caption{Radial functions as obtained from diagonalization of finite difference version of Eqs. (\ref{era1}) and  (\ref{era2}) 
for  $m=0$, $\eta=1$ and $N=200$ mesh points, radius of the flake $R=80$ nm and   zigzag boundary condition $f_2(r=R)=0$.
 Central, three-point finite difference quotient for the derivative is used 
with the exception of the last point on a mesh $N$, where a backward quotient is applied. 
The results for the eigenvalues  $n'=1,2,3,4$  of Table \ref{pozytywne} are plotted in (a,b,c) and (d), respectively.
(a,c) correspond to the spurious solutions, and (b,d) to the regular ones.
The eigenfunctions are determined up to a normalization constant so the arbitrary units were used.
} \label{ffdiag}
\end{figure}

The wave functions for $n'$ from 1 to $4$ are displayed in Fig. \ref{ffdiag}. With the applied boundary conditions $f_1$ is real and $f_2$ is imaginary.
The spurious states [Fig. \ref{ffdiag}(a,c)] correspond to the real part of $f_1$ and imaginary part of $f_2$ that rapidly oscillate between the  nearest-neighbor mesh points. The solution given by  Fig. \ref{ffdiag}(b,d) corresponds to the analytical eigenstates. 

\begin{figure}[htbp]
\includegraphics[width=0.8\columnwidth]{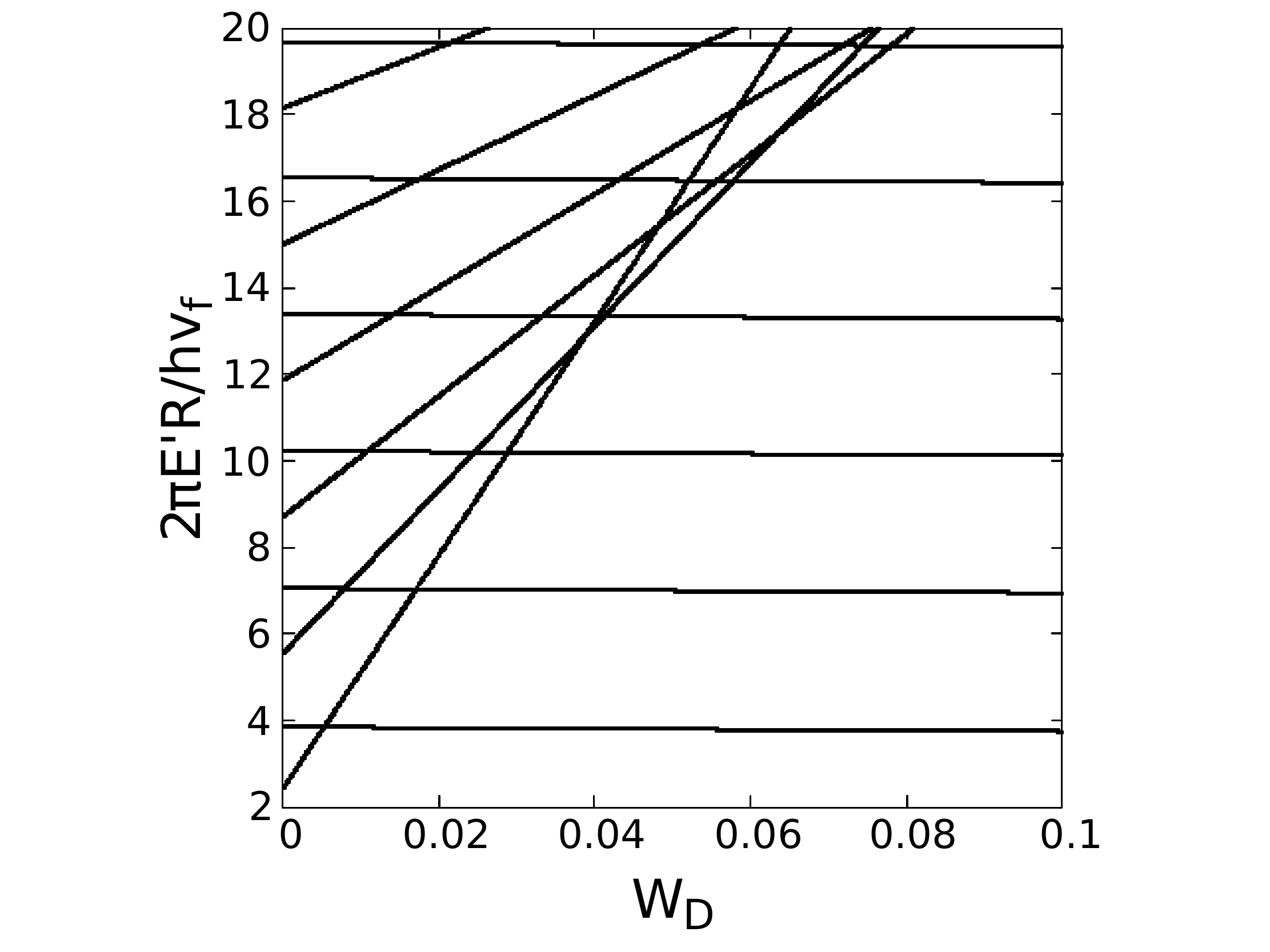}
\caption{Energy levels for $\eta=1$ and $m=0$ shifted by the Wilson term that separates the spurious energy levels from the original ones -- see Table \ref{pozytywne}.
Results were obtained for $N=200$ mesh points.
} \label{ffdia}
\end{figure}

In the spurious solutions $\langle {\bf k}^2 \rangle$ is large due to the rapid wave function oscillations (see Fig. \ref{ffdiag}(a,c)).
Using this fact one can numerically remove the spurious states from the low-energy part of the spectrum. The procedure  \cite{stacey,tanimura,wilson,susskind}
involves an extra artificial energy operator of form 
\begin{equation}
H_D=-W_D \hbar  v_f  \nabla ^2 \sigma_z dr,
\end{equation}
where $W_D$ is a dimensionless Wilson parameter. Figure \ref{ffdia} shows the energies of the $m=0$, $\eta=1$ states
(same as in Fig. \ref{ffdiag} and Table \ref{pozytywne}), where $H_D$  is introduced by the first order perturbation
$E_{mn,\eta}'=E_{mn,\eta}+\langle \Psi_{m,n,\eta}| H_D |\Psi_{m,n,\eta}\rangle $. The states identified as spurious in the context of Table \ref{pozytywne}
and Fig. \ref{ffdiag} are indeed removed from the low-energy spectrum, while the actual solutions are only weekly
affected by the Wilson term.
\begin{table}
\begin{tabular} {c|ccccc}
&$n'=1$ & $n'=2$ &  $n'=3$  & $n'=4$ & $n'=5$ \\  \hline
$N=200$ & {\bf 2.419} & 3.850 & {\bf 5.548} & 7.049 & {\bf 8.696} \\
$N=400$ & {\bf 2.412 }& 3.841 & {\bf 5.535} & 7.033 & {\bf 8.675} \\
$N=800$ & {\bf 2.409 } & 3.836 &{\bf 5.528} & 7.024 & {\bf 8.665} 
\end{tabular}
\caption{The eigenergies obtained by diagonalization of the system of  Eqs. (\ref{era1}) and  (\ref{era2}) with the finite
difference approach for $m=0$ and $\eta=1$. The eigenergies of the spurious states that are put in bold,
correspond to the regular solutions but obtained for a different $m$ -- cf. Table \ref{besle}, which results in the fermion doubling. } \label{pozytywne}
\end{table}

\section{The mesh sweeping method}

We are ready to introduce the method which is the purpose of this work. The method provides the solution of the Dirac equation on a finite difference mesh and is free of the spurious states,
so  no discrimination of the rapidly oscillating states is necessary. The general system of equations given by Eq. (\ref{e1})
is transformed to a form that allows for sweeping the mesh from the edge of the flake to its center. For this purpose the radial derivative
is replaced by a two-point finite difference quotient $f'(r)=\frac{f(r)-f(r-dr)}{dr}$. The sweeping finite difference scheme is \begin{widetext}
\begin{eqnarray}
f_1(r-dr)&=&\frac{idr}{\hbar} \left[\left(U_B(r)-E\right)\frac{f_2(r)}{v_f}+\left(-\frac{i\hbar}{dr}+\eta \frac{ieBr}{2}+\frac{i\hbar\eta m}{r}\right) f_1(r)\right], \label{chc} \\
f_2(r-dr)&=&\frac{idr}{\hbar} \left[\left(U_A(r)-E\right)\frac{f_1(r)}{v_f}-\left(\frac{i\hbar}{dr}+\eta \frac{ieBr}{2}+\frac{i\hbar\eta (m+\eta)}{r}\right) f_2(r)\right] \label{chc2}
\end{eqnarray} \end{widetext}

\begin{figure}[htbp]

\includegraphics[width=\columnwidth]{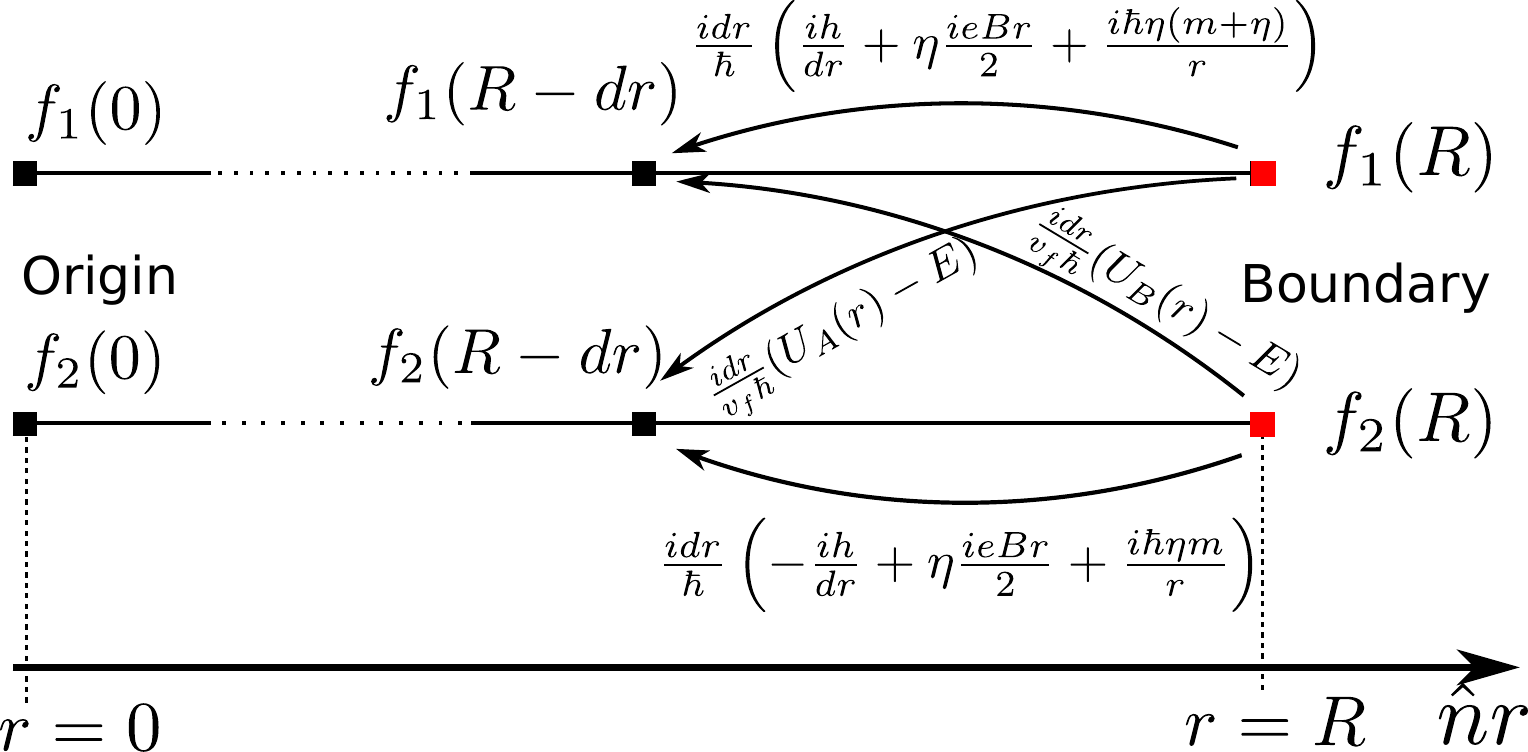} 
\caption{The schematics of the sweeping method. The boundary conditions $f_1(R)$, $f_2(R)$ are applied
at the external edge of the flake. Eqs. (\ref{chc}) and (\ref{chc2}) allow for evaluation of the wave function components to the left for the energy $E$ which is a parameter to be determined by the boundary condition to be reproduced at the origin. } \label{t2}
\end{figure}

\begin{figure}[htbp]
\begin{tabular} {c}
(a) \includegraphics[width=0.6\columnwidth]{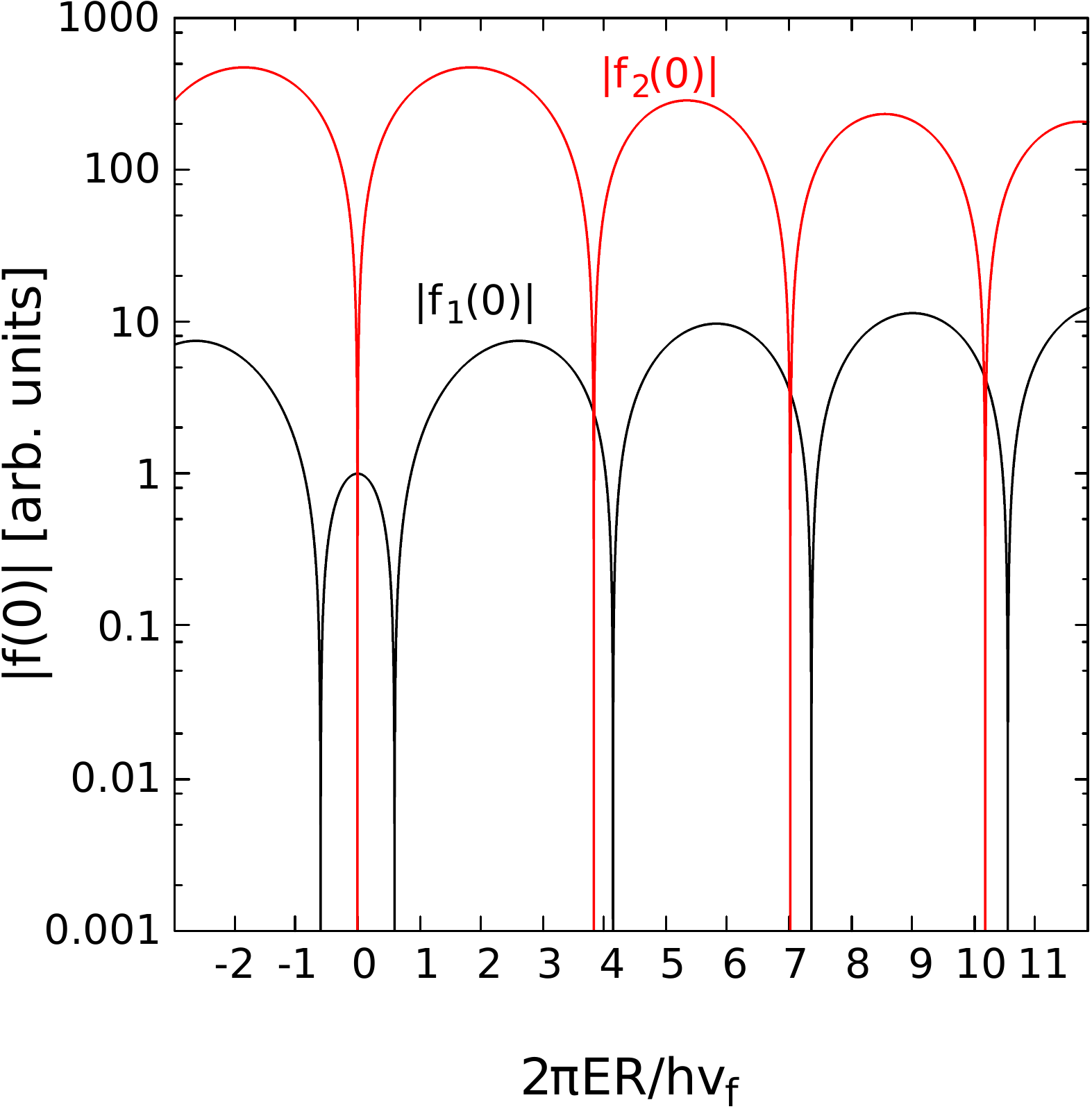} \\ (b) \includegraphics[width=0.6\columnwidth]{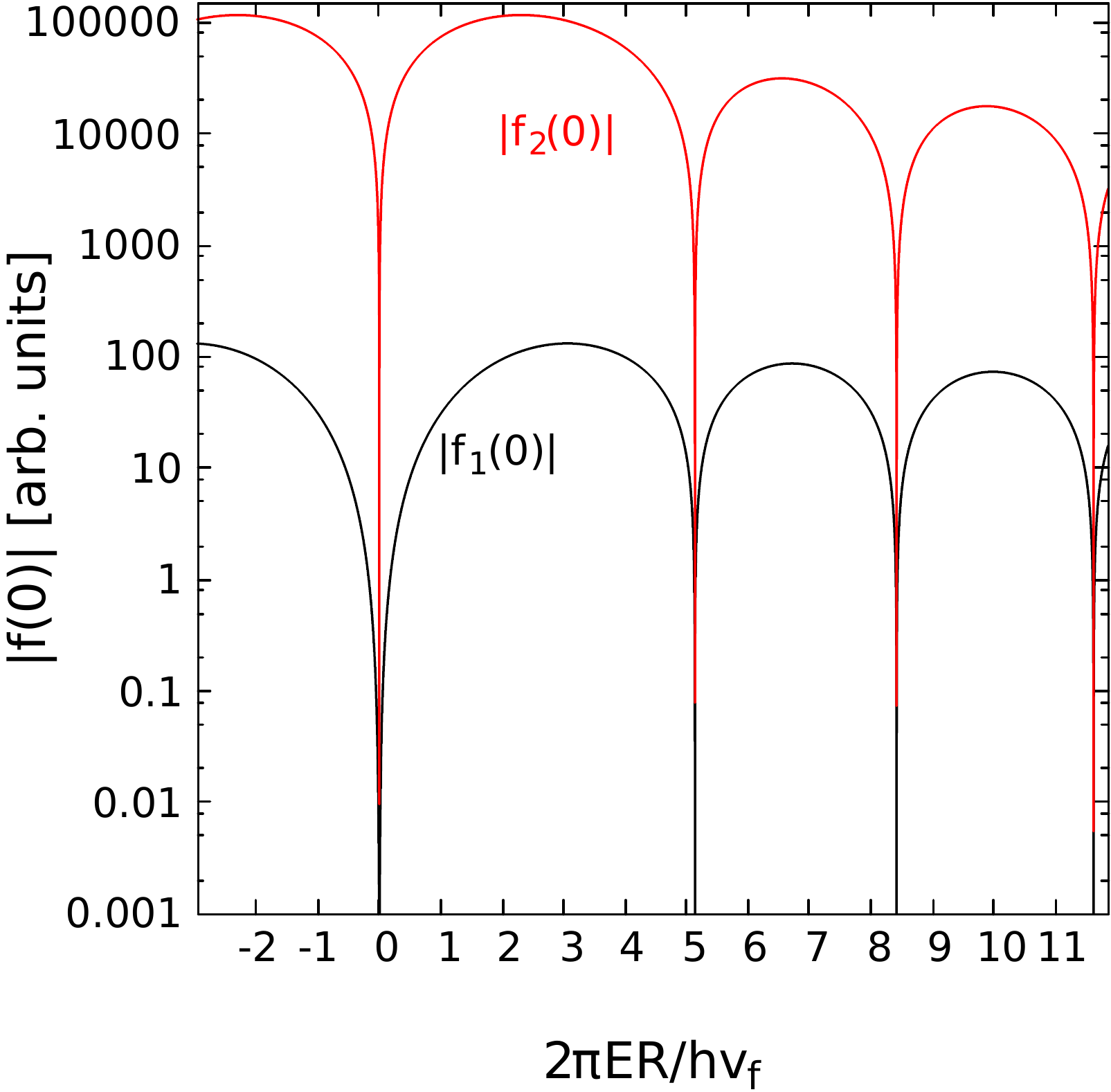}\\
(c) \includegraphics[width=0.6\columnwidth]{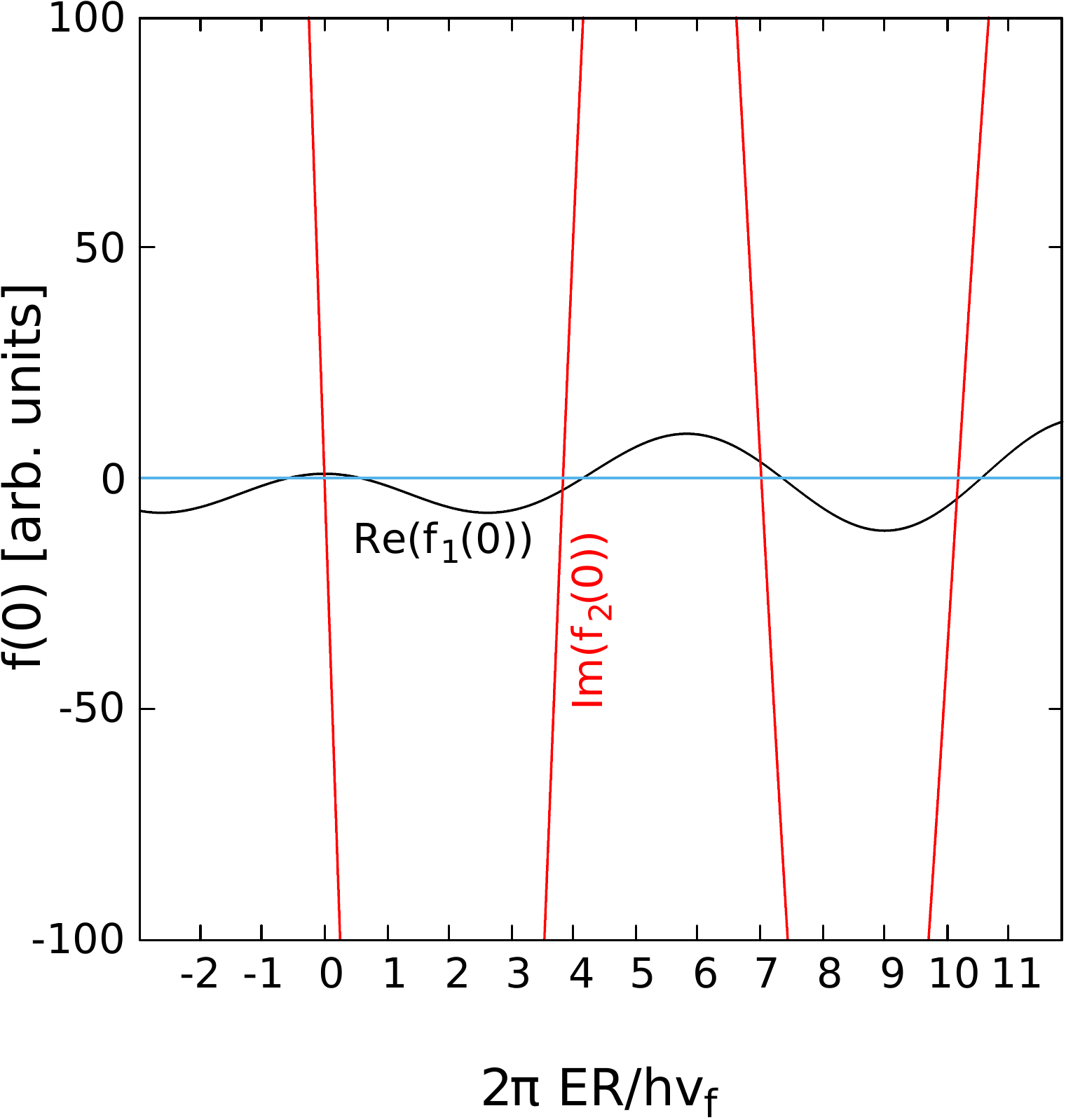} \\(d) \includegraphics[width=0.6\columnwidth]{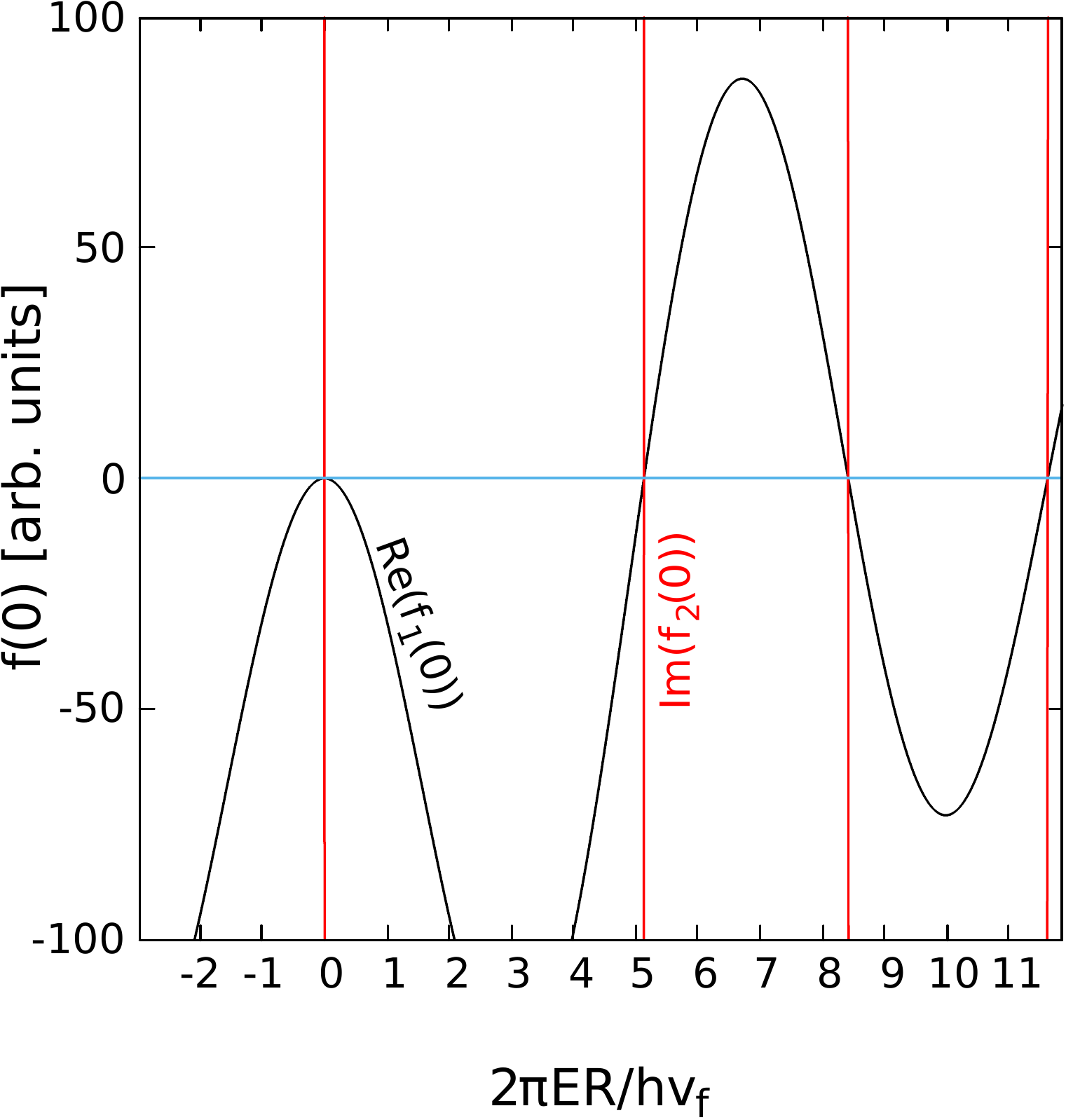}
\end{tabular}
\caption{Absolute values (a,b), real and imaginary parts of the radial functions $f_1$ and $f_2$ (c,d) at the origin
for the sweeping method starting at the edge of the flake for the $K$ valley $\eta=1$. Panels (a,c) were calculated for $m=0$ and (b,d) for $m=1$.
} \label{center}
\end{figure}

The schematics of the calculations given in Fig. \ref{t2}.
The procedure starts at the edge of the flake $r=R$, where the boundary conditions are applied. For the zigzag boundary condition we set
$f_2(R)=0$ and $f_1(R)=1$. The eigenfunctions can be normalized after the entire procedure. 
With the starting values of $f_1$ and $f_2$ at $r=R$ one can proceed to the center of the flake using Eqs. (\ref{chc}) and (\ref{chc2}).
At the origin $f_1$ and $f_2$ need to vanish when $m \neq 0 $ and $m+\eta \neq 0$, respectively. That, in turn is realized only for discrete values of the energy. For $\eta=1$, the values of the components of the wave function for $r=0$ are given  in Fig. \ref{center} for $m=0$ (left column) and $m=1$ (right column). For $m=0$ the component $f_2$ needs to vanish at the origin, since 
the angular momentum quantum number corresponding  to the second component is nonzero, $m+\eta=1$. 
The position  of the energy eigenvalues can be found as the minima of the absolute values of the $f_2(r=0)$ [Fig. \ref{center}(a)] or equivalently, by zeroes of the imaginary part of $f_2(r=0)$  [see Fig. \ref{center}(c)] as a function of the energy.
The minima [Fig. \ref{center}(a)] and the zeroes [Fig. \ref{center}(c)]  correspond to the actual eigenvalues and the spurious solutions are missing [cf. Tables \ref{besle} and \ref{pozytywne}]. 
For $m=1$ both $f_1$ and $f_2$ components need to vanish at the origin, and indeed no shift of the minima of Fig. \ref{center}(b) or zeroes of Fig. \ref{center}(d) is observed on the energy scale.
Besides the solutions of the Bessel form,  the results of Fig. \ref{center} indicate the presence of the zero-energy levels that are supported by the zigzag edge \cite{gru}.

\begin{figure}[htbp]
 \includegraphics[width=0.8\columnwidth]{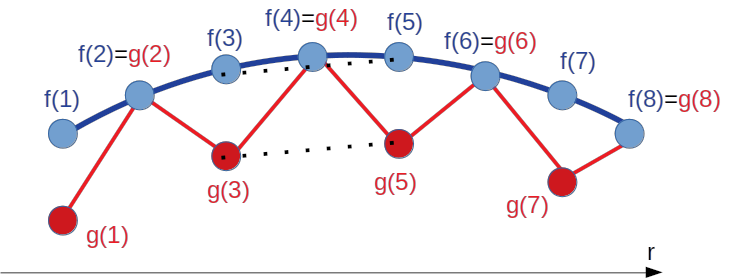} 
\caption{A schematic drawing of a smooth solution $f$ given on a radial mesh, and a
spurious one $g$, rapidly oscillating from one lattice point to another. 
The central quotient for the derivative for point (4) can be equal for both solutions $f'(4)\simeq\frac{f(5)-f(3)}{2dr}=\frac{g(5)-g(3)}{2dr}\simeq g'(5)$ (see the black dotted lines). The present sweeping method is based on a asymmetric formula for which the  quotients are very different for both functions $\frac{f(5)-f(4)}{dr}\neq \frac{g(5)-g(4)}{dr}$, so that the spurious solution is associated with a very different energy.
} \label{spur}
\end{figure}

The success of the sweeping approach is due to the two-point derivative that does not allow for saw-like oscillations of the wave function. To explain this let us look at Fig. \ref{spur} where two
functions: a smooth one $f$ and a rapidly oscillating one $g$ are plotted on a radial mesh.
For even indexes functions $f$ and $g$ are equal and for odd indexes a shift between the solutions 
is present. From the point of view of the central quotient of the first derivative both the solutions
are identical; in each mesh point the same value of the quotient are obtained, and there is no
relation between the odd and even indexes on the mesh points, which allows for the spurious states to appear in the low-energy part of the spectrum. The Wilson term given by Eq. (10) introduces the 
second derivative with the quotient $f''(i)\simeq \frac{f(i+1)+f(i-1)-2f(i)}{dr^2}$ that links the odd and even points and removes the rapidly oscillating states from the low-energy spectrum. 
The sweeping method that uses the backward quotient $f'(i)=\frac{f(i)-f(i-1)}{dr}$ and passes through each mesh point consecutively introduces the link between the even and odd mesh points.
The two-point formula for the smooth and rapidly oscillating solutions produce very different values of the quotient.  
 
Tabel \ref{precision} shows that the convergence of the results of the present method 
to the exact solution is linear as a function of $dr$ (or $1/N$). Although the
convergence is slow, the numerical cost of the calculation increases only linearily with $N$,
and the method needs to keep track of only four complex numbers for the wave function components
at adjacent mesh points, so one can approach the exact solution arbitrarily close at a negligible numerical cost.

\begin{table}
\begin{tabular} {c|ccccc}
$N$&  $E(n=1)$ & $E(n=2)$ &  $E(n=3)$  \\ \hline
 100 & 3.853094 &  7.081613 &   10.304313 \\ 
 200 & 3.842100 & 7.046336    &   10.231384 \\ 
 400 & 3.836828 & 7.030400&    10.200578 \\ 
 800 & 3.8342475 & 7.022854&     10.186564\\ 
 1600 & 3.832970 & 7.019184&    10.179904\\ 
 3200 & 3.832339 & 7.017377& 10.176656 \\ 
 6400 & 3.832023 & 7.016478& 10.175054 \\ \hline
 exact & 3.831706 & 7.015587 &  10.173468 \\ \hline\hline
$N$&  $\Delta E(n=1)$ & $\Delta E(n=2)$ &  $\Delta E(n=3)$  \\ \hline 
100& 0.021388&	0.066026	&0.130845 \\
200& 0.010394	&0.030749	&0.057916\\
400&0.005122	&0.014813	&0.027110\\
800&0.002541	&0.007267	&0.013096\\
1600&0.001264	&0.003597	&0.006436\\
3200&0.000633	&0.00179	&0.003188\\
6400 & 0.000317	&0.000891	&0.001586\\ \hline
\end{tabular}
\caption{First seven rows show the three lowest positive energy levels for $m=0$ and $\eta=1$ in the units of $\frac{R}{v_f \hbar}$ 
as calculated with the sweeping method for $U_A=U_B=0$ and $B=0$ as a function of the number of mesh points $N$. 
The central row gives the exact results (cf. Table \ref{besle}).
Bottom part of the table: rows 10th and below show the difference of the numerical and exact results as function of $N$.
} \label{precision}
\end{table}

\begin{figure}[htbp]
\begin{tabular} {l}

(a) \includegraphics[width=0.6\columnwidth]{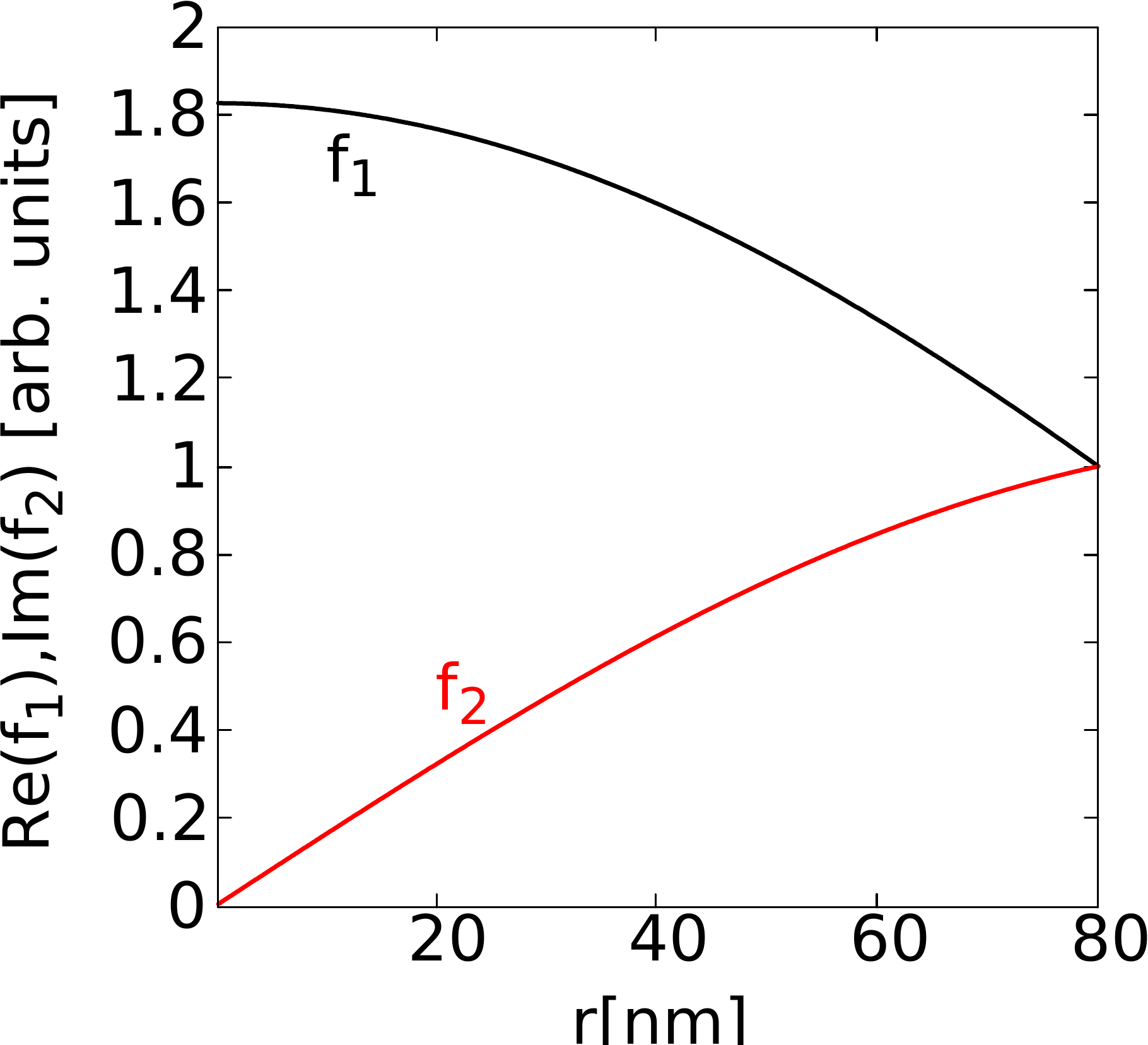} \\
(b) \includegraphics[width=0.6\columnwidth]{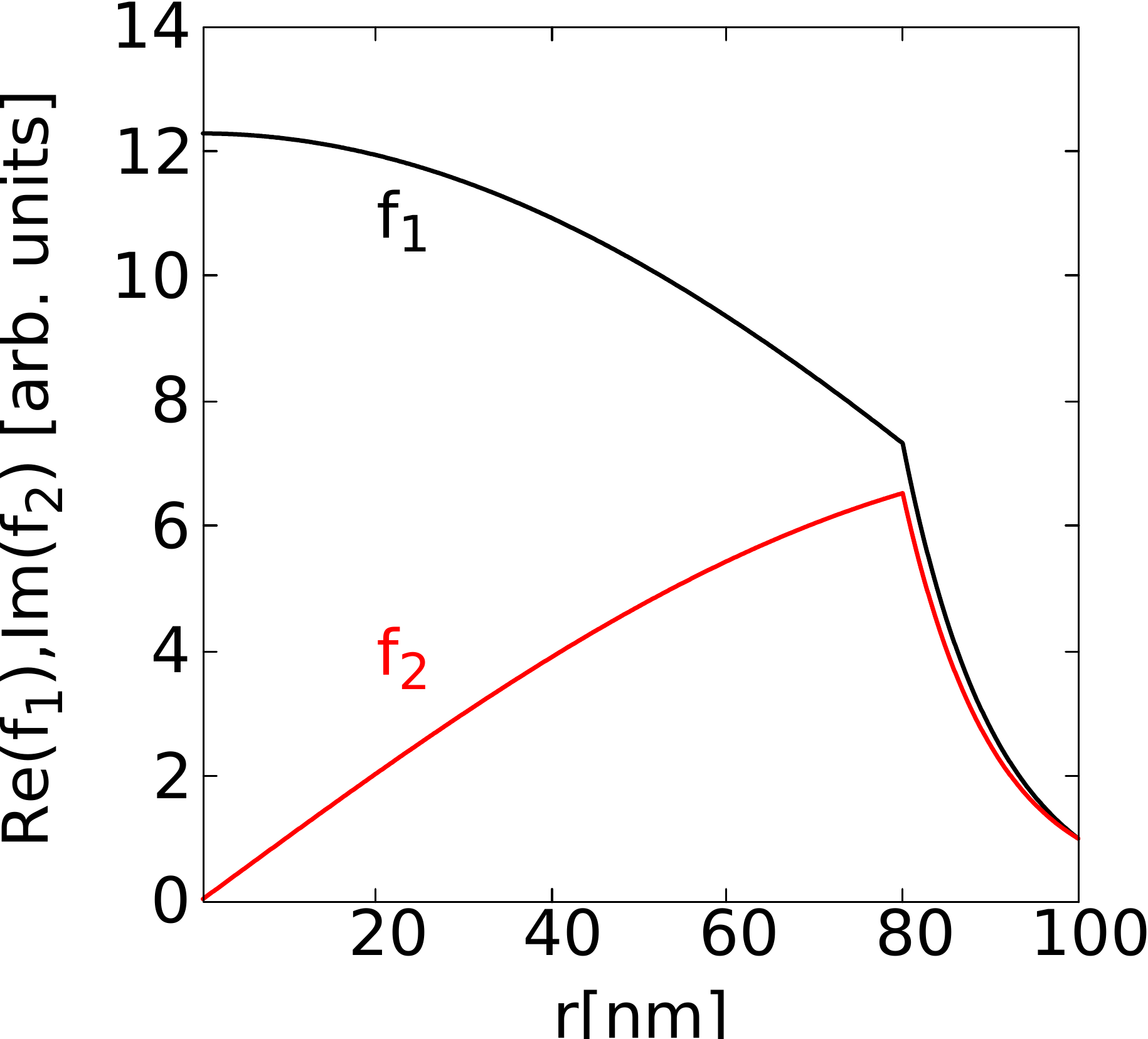} \\
(c) \includegraphics[width=0.6\columnwidth]{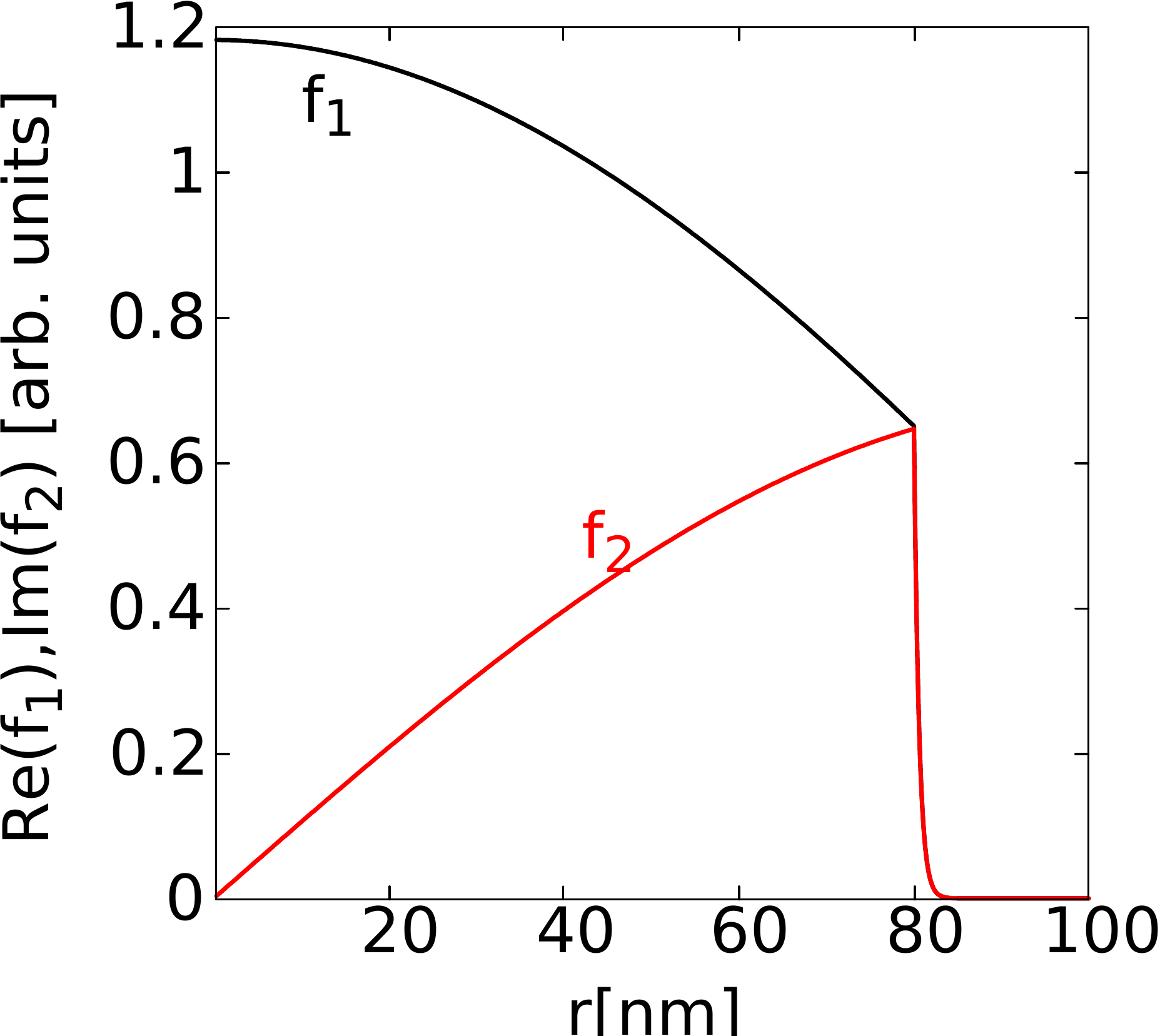} 
\end{tabular}
\caption{Real and imaginary parts of the $f_1$ and $f_2$ radial functions respectively for 
an infinite mass boundary condition at the edge of the flake $R=80$ nm (a) and $R=100$ nm (b,c).
We consider $\eta=1$ state with $m=0$. In (a) no external potential is applied. In (b) and (c) the external potential is introduced $U_A=-U_B$ with $U_A=0.1$ eV (b) and $U_A=1$ eV (c).} \label{infma}
\end{figure}

\section{Examples of applications}
The approach is suitable
for any radial problems involving the effective Hamiltonian that stems from the Dirac equation for a given valley. 
This section provides examples of applications.

\subsection{Limit of infinite-mass confinement}
In the precedent section the zigzag-boundary conditions were used that result from a specific  termination of the crystal lattice. 
The confinement of the particle described by the Dirac Hamiltonian can be induced provided by e.g. a spatial dependence
of the energy gap \cite{nori} of equivalently by an external potential due to e.g. a substrate   \cite{sachs}. 

For $U_A\neq U_B$ an energy 
gap is opened in the dispersion relation near the Dirac points and the carriers acquire finite masses. 
For $U_A$ and $U_B$ that diverge to $\infty$ and $-\infty$ when $r>R$, respectively, the mass in the outside of the confinement is infinite and the wave functions of finite energies vanish at  $r>R$ \cite{berry,huang,nipre}.
The boundary condition  for this type of confinement is derived from vanishing current at the edge \cite{berry} of the confinement area.
The probability density current that is derived from the Hamiltonian \cite{berry} is 
${\bf j}=2v_F\left[\Re(\Psi_1^* \Psi_2),\eta \Im(\Psi_1^*\Psi_2)\right]$. 
For the infinite mass at the outside of the dot, the orthogonal component of the current must vanish $(\cos\phi,\sin\phi)\cdot {\bf j}=0$, which implies $\tan\phi=-\eta \frac{\Re(\Psi_1^* \Psi_2)}{\Im(\Psi_1^*\Psi_2)}$.

With Eq. (2) this condition is translated for the radial functions 
as $\tan\phi=-\eta \frac{\Re(f_1^*(R) f_2(R) \exp(i\eta\phi))}{\Im (f_1^*(R) f_2(R) \exp(i\eta \phi))}.$
To fulfill this condition we set at the end of the flake $f_1(R)=1$, and $f_2(R)=i$, for which 
one gets $\tan\phi=\eta \tan \eta \phi$, which is fulfilled for both the valleys $\eta=\pm 1$. 

Let us look at the appearance of the wave functions confinement by the mass boundary.
We consider a spatial dependence of the energy gap as introduced by finite $U_A=-U_B$  potentials.  
Figure \ref{infma} shows the real part of $f_1$ and the imaginary part of $f_2$ for $\eta=1$ and $m=0$. 
At the end of the flake the infinite mass boundary condition is applied. The components
$f_1$ and $f_2$ are -- as for the zigzag boundary --  purely real and purely imaginary, respectively. Moreover, the infinite-mass  boundary condition implies $\Re\{f_1(r=R)\}=\Im\{f_2(r=R)\}$.
The results of Fig. \ref{infma}(a) were obtained for a flake of radius $R=80$ nm in the absence of the external potentials.
In Fig. \ref{infma}(b,c) we extended the flake to $R=100$ nm but introduced the energy gap beyond $r>R'=80$ nm 
with $U_A=-U_B$. In Fig. \ref{infma}(a), $U_A=0.1$ eV and $U_A=1$ eV in Fig. \ref{infma}(c).
We can see that for larger $U_A$ the functions penetrate only weakly into the gapped region for $r>R'=80$ nm, and the values
$\Re\{f_1\}$ and $\Im\{f_2\}$ become equal at $r=R'$,  where the gap is introduced. In this way the infinite boundary condition
is found in the limit of large $U_A$ at $r=R'$. 

\subsection{The quantum ring spectra}
The method with a slight modification can be applied to a problem of a graphene quantum ring \cite{tho}. 
From the flake of the radius of $R=80$ nm we remove the central disk of a radius of $R_i=40$ nm.
The infinite mass boundary condition in the inner edge of the flake reads $f_2(R_i)/f_1(R_i)=-i$. 
We start from the external edge $R$ as in the precedent subsection. The sweep at the finite difference 
mesh stops ar $R_i$. From the inner boundary condition we construct a function $F(E)=f_2(R_i;E)+if_1(R_i;E)$
and we look for its zeroes as a function of the energy. 

\begin{figure}
\begin{tabular}{cc}
(a) \includegraphics[width=.45\columnwidth]{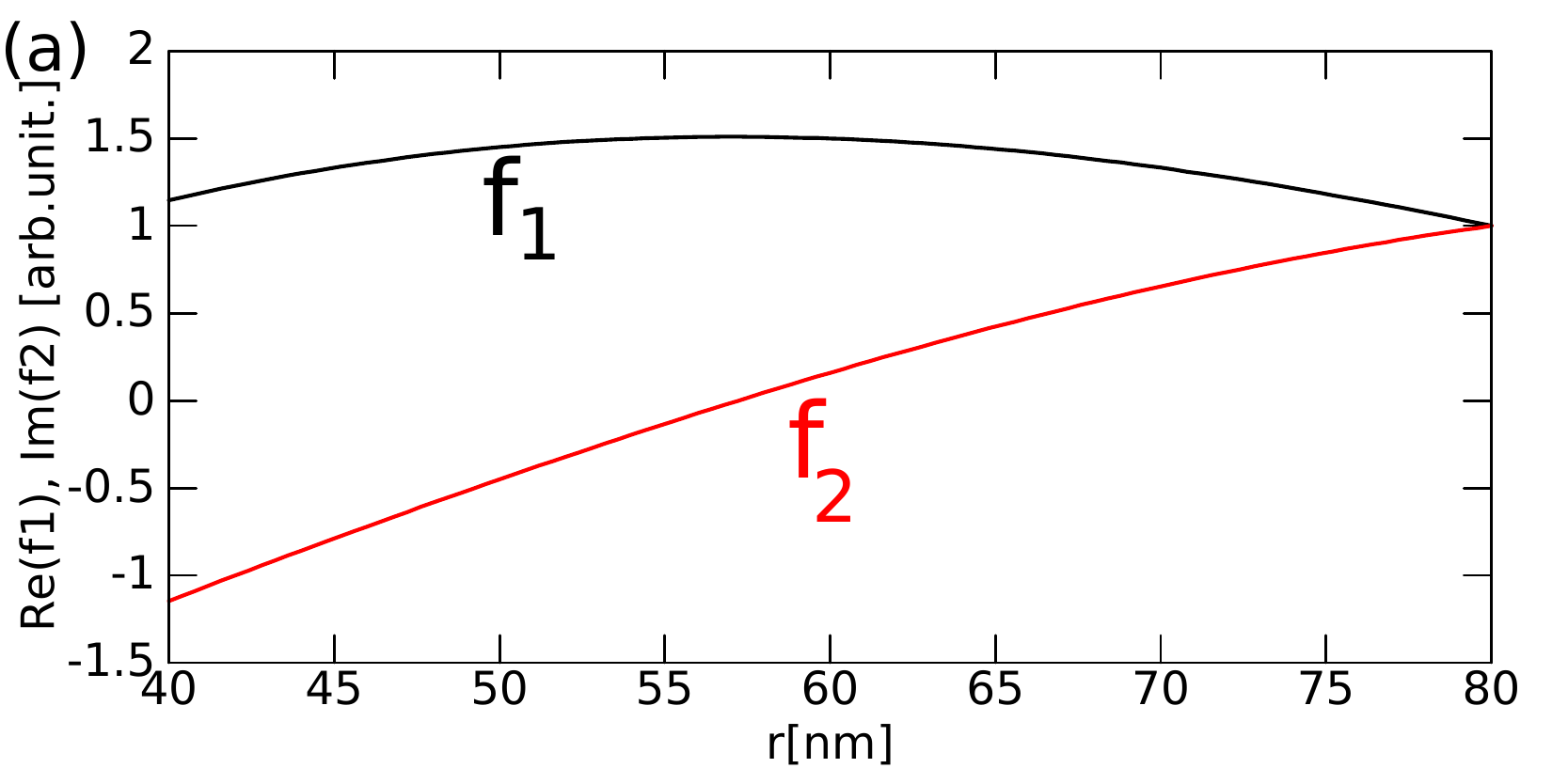}   &
(b) \includegraphics[width=.45\columnwidth]{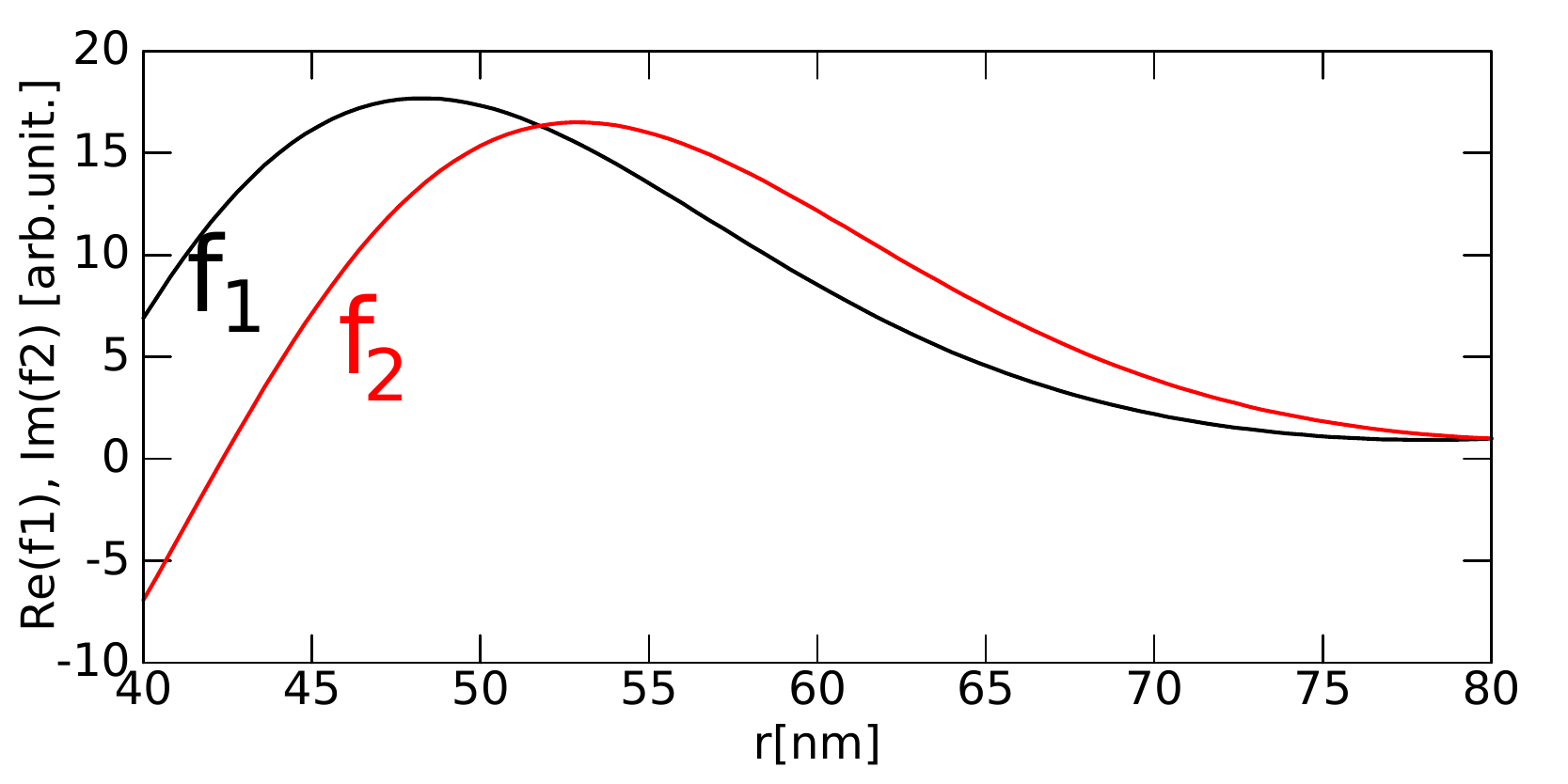}  \\ 
(c) 
 \includegraphics[width=.45\columnwidth]{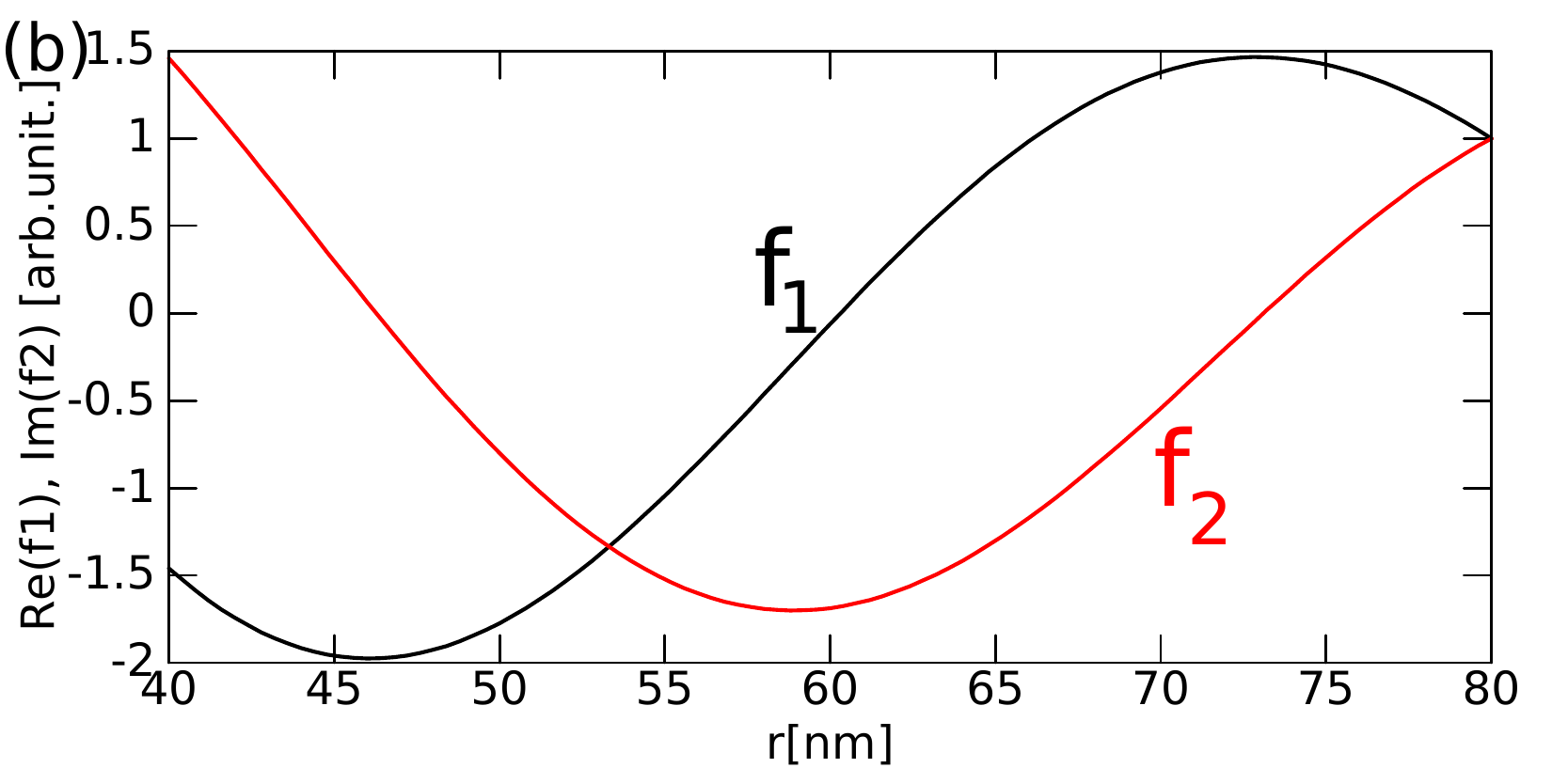}  &
(d)  \includegraphics[width=.45\columnwidth]{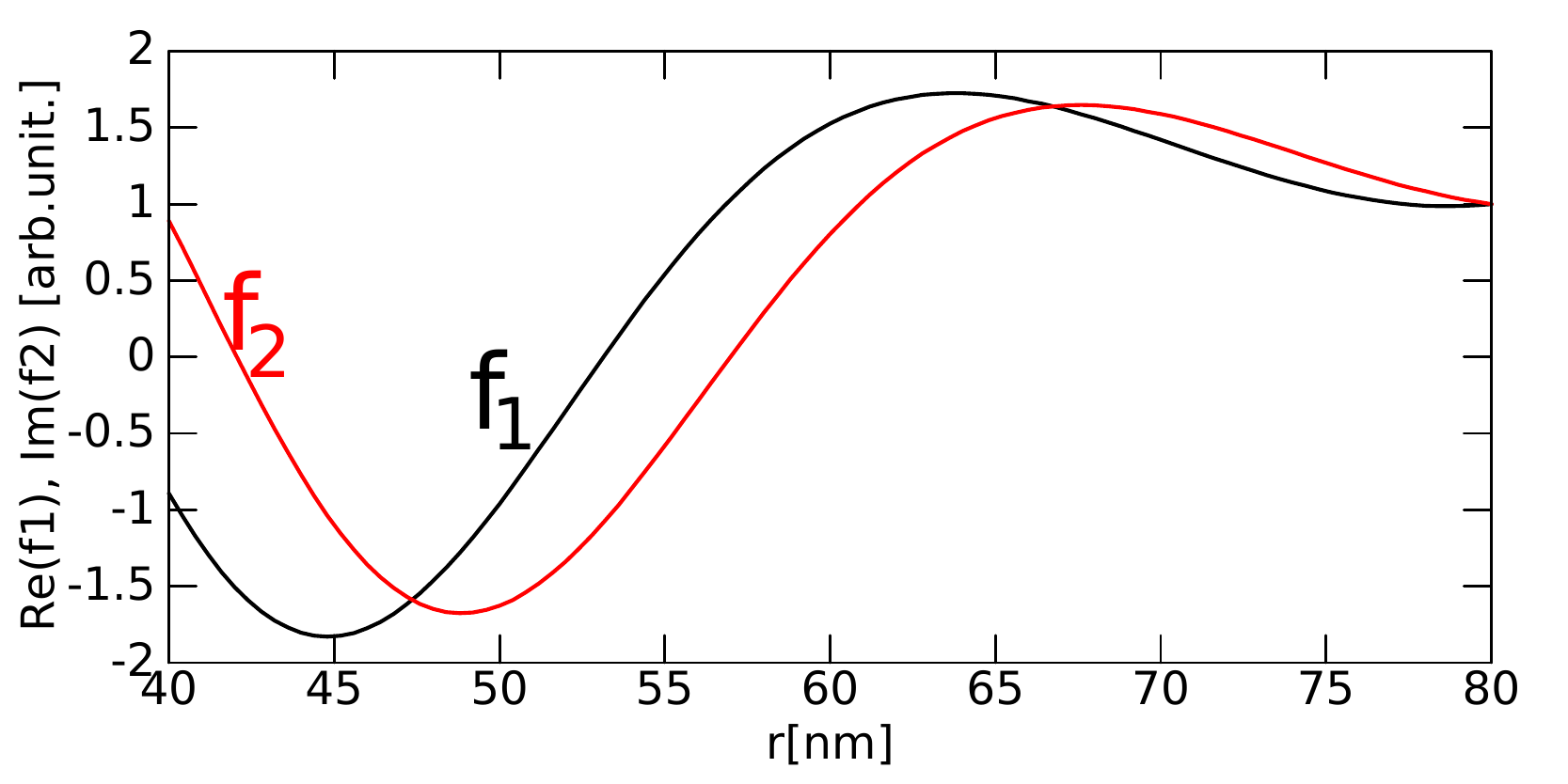}  \end{tabular}
\caption{The radial functions with $m=0$ for the $K$ valley ($\eta=1)$ of the (a,b) $n=+1$, (c,d) $n$=+2 state in a ring with inner (outer) radius $R_i=40$ nm ($R=80$ nm) 
and the infinite mass boundary conditions and $B=0$ (a,c), $B=5$ T (b,d).
}  \label{bbb}
\end{figure}

\begin{figure}
 \includegraphics[width=\columnwidth]{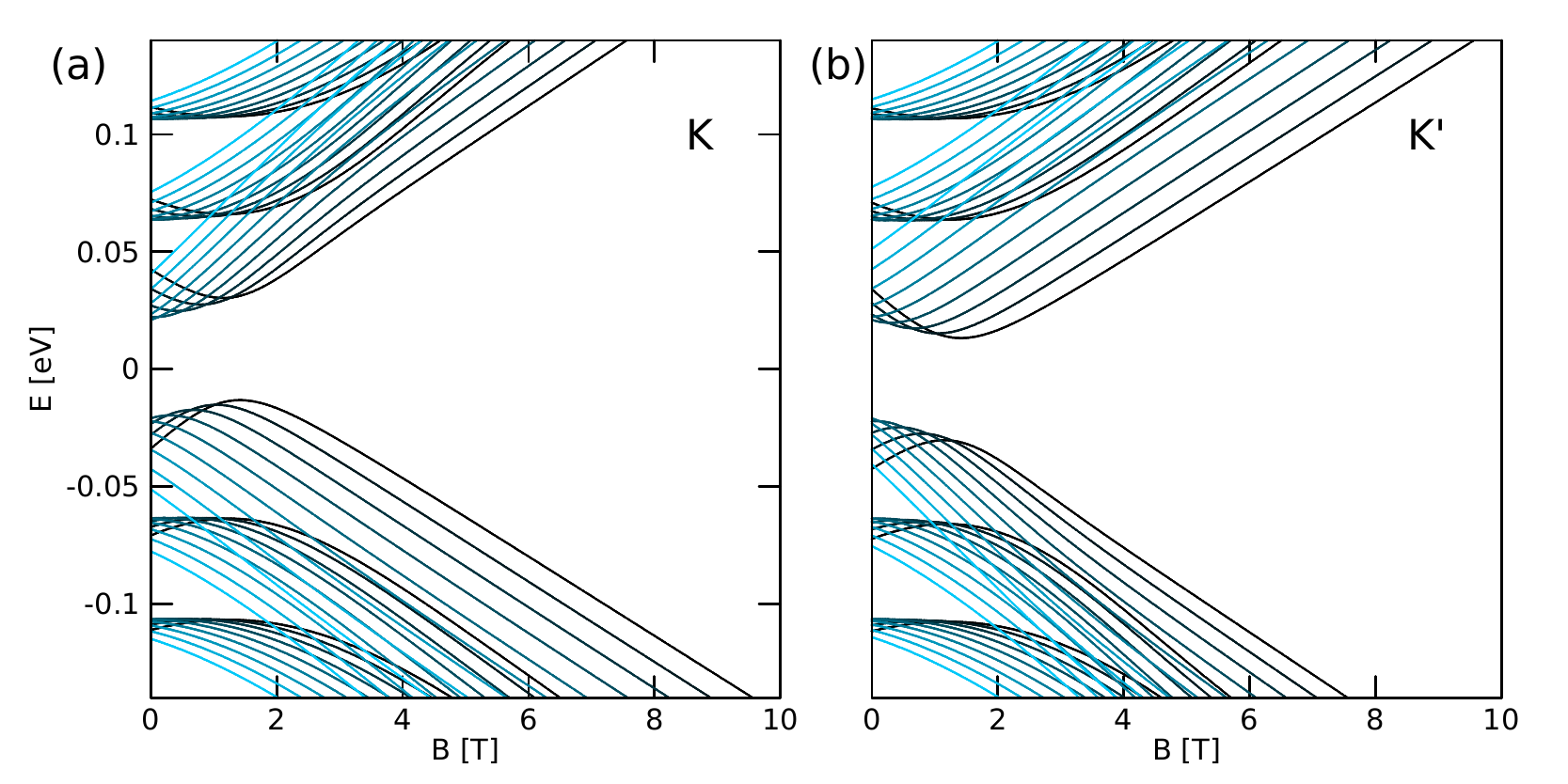} 
\caption{The energy levels of a ring with inner (outer) radius $R_i=40$ nm ($R=80$ nm) and with infinite mass boundary conditions, with $m=0,\pm 1,...,\pm 4$. The black curve shows the $m=-4$ level, and the brighter curves have higher $m$.
} \label{acac}
\end{figure}

The $m=0$ wave functions for the $K$ valley ($\eta=1$) found in this way are displayed in Fig. \ref{bbb} for $n=1$ (a,b) and $n=2$ (c,d).
The energy spectra for $\eta=\pm 1$ are given in Fig. \ref{acac} for $|m|\leq 4$. In the lowest energy states of the conduction and the valence
bands one observes the characteristic angular momentum transitions with the Aharonov-Bohm periodicity of $\Delta B\simeq 0.385$ T \cite{pers}, that 
corresponds to a flux quantum threading a ring of an effective radius $59$ nm -- close to the central radius of the ring $R_c=(R+R_i)/2$.

\subsection{Comparison with analytical results in the external magnetic field}
\begin{figure}
 \includegraphics[width=\columnwidth]{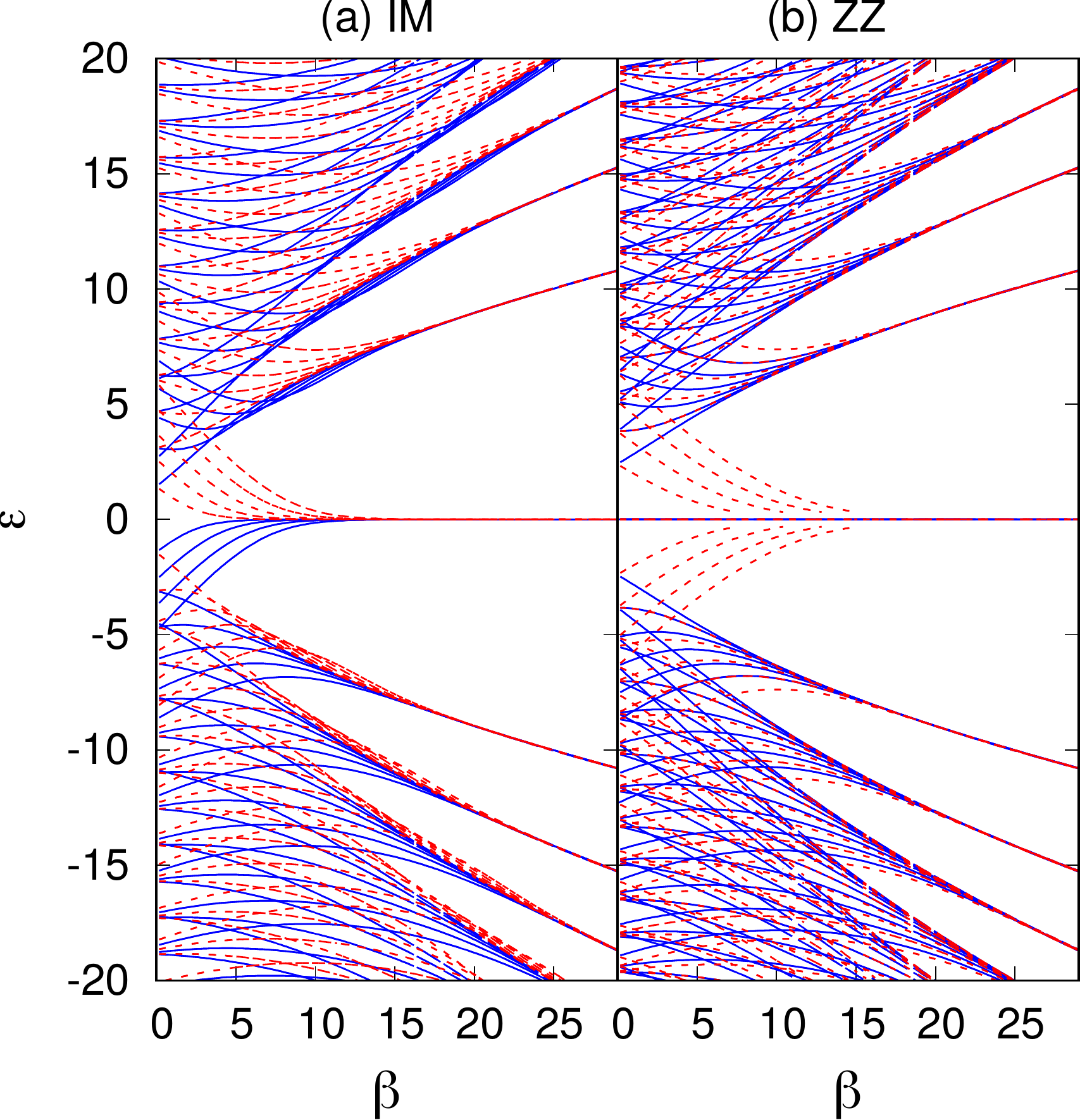} 
\caption{The energy spectrum calculated for graphene quantum dot of radius $R=70$ nm
with the present approach, to be compared with Fig. 1 of Ref. \cite{gru}. 
In (a) the infinite mass boundary condition is applied and in (b) the zigzag boundary.
The dashed lines correspond to $K'$ valley and the solid ones with the $K$ valley.
The energies and the magnetic field are expressed in dimensionless units, $\epsilon=\frac{ER}{hv_F}$ 
and $\beta=\frac{eBR^2}{2\hbar}$. Energy levels for $m\in[-4,4]$ are displayed.
} \label{grur}
\end{figure}

In the absence of the external potential $U_A=U_B=0$, or for a massless particle, analytical results were obtained for the Weyl equation in the external magnetic field in Ref. \cite{gru} for both zigzag and infinite mass boundary conditions. We adopted the graphene parameters of this work \cite{gru} $t=2.7$ eV, the lattice constant $0.142$ nm and the radius of the dot $R=70$ nm. The results for the 500 mesh points are displayed in Fig. \ref{grur}
with a perfect agreement with Fig. 1 of Ref. \cite{gru}.

\section{Summary}

We have presented the finite difference method, applicable to Dirac carriers in circular confinement,
that sweeps the mesh from the external edge of the system, where the boundary conditions are defined, to the center. The energies of the confined states are pinned by the
internal boundary conditions. The method is simple, free of spurious solutions and the fermion doubling problem.
Applications to circular flakes, quantum rings, and the energy gap modulated in space were presented.

\section*{Acknowledgments}
This work was supported by the National Science Centre (NCN) according to decision DEC-2016/23/B/ST3/00821.

\bibliographystyle{elsarticle-harv}



\end{document}